# Poisson-Boltzmann theory with non-linear ion correlations

Mao Su[1,2,3], Zhijie Xu[3] and Yanting Wang[1,2]*

[1]CAS Key Laboratory of Theoretical Physics, Institute of Theoretical Physics, Chinese Academy of Sciences, 55 East Zhongguancun Road, P. O. Box 2735, Beijing 100190, China

[2]School of Physical Sciences, University of Chinese Academy of Sciences, 19A Yuquan Road, Beijing 100049, China

[3]Computational Mathematics Group, Pacific Northwest National Laboratory, Richland, Washington 99352, USA

*E-mail: wangyt@itp.ac.cn

**Abstract**

The Poisson-Boltzmann (PB) theory is widely used to depict ionic systems. As a mean-field theory, the PB theory neglects the correlation effect in the ionic atmosphere and leads to deviations from experimental results as the concentration or charge valance increases. A modified PB theory including ion correlation effect while retaining its simplicity is critical for many important applications in which ion correlation effect can be significant. In this paper, we present a new model to incorporate ion correlations into the original PB equation by utilizing the Green's function with a non-linear form of the self-energy, which is different from the linear self-energy equation obtained by the Field-Theoretic (FT) approach. Both equations are solved numerically and compared with our molecular dynamics (MD) simulation. The co-ion distribution calculated by the FT approach deviates significantly from the MD simulation, while our results for both counter-ion and co-ion distributions are justified by the MD simulation.

## 1. Introduction

Reliable theoretical models for ionic systems are of great fundamental importance for basic research, especially for the study of soft-matter [1, 2]. As a mean-field theory, the Poisson-Boltzmann (PB) model developed by Debye and Hückel [3] grasps the main features of ionic systems. By using the average electrostatic potential instead of the potential of mean force [4], the ion density $\rho$, which approximately satisfies the Boltzmann distribution, along with the electrostatic potential $\Phi$, should satisfy the Poisson's equation, leading to the original PB equation as

$$-\epsilon \nabla^2 \Phi = \sum_i e z_i \rho_{si} \exp(-e z_i \Phi / k_B T), \qquad (1)$$

where $\epsilon$ is the dielectric constant, $e$ is the unit charge, $z_i$ and $\rho_{si}$ are the valance and bulk density of the $i$th ion species, respectively, $k_B$ is the Boltzmann constant, and $T$ is the temperature. The PB equation is often used to solve the electrostatic potential around a molecule or a certain boundary, as well as to obtain physical properties such as ion distribution, solvation free energy, and activity coefficient with the knowledge of statistical physics.

The PB theory has many applications in various research fields, ranging from investigating the electrostatic properties of charged molecules and ionic solutions [5] to studying the structures and flexibilities of membranes [6, 7], as well as calculating charge distributions on the surface of biomolecules such as DNA, RNA and proteins [8, 9]. In its application, it is well known that the PB theory will qualitatively fail when ion correlations become significant [10-13], so a variety of theoretical models, whether based on PB theory or not, have been developed to incorporate the ion correlation effect. The integral equation method with the Hypernetted Chain (HNC) approximation is known to be very accurate for ionic systems [14, 15], but theoretically the physical picture behind this approximation is vague, and practically the HNC equations can be hard or even impossible to solve in certain cases [13, 16]. Other methods like the density functional theory [17, 18] and the modified Poisson-Boltzmann equation developed by Outhwaite and Bhuiyan [19] involve elaborate mathematical expressions that are too complex for practical applications. Therefore, some exact mean-field theories are developed to yield accurate results while keeping their mathematical expressions as simple as the Debye-Hückel (DH) equation, which is a linearized instance of the PB equation, written as

$$\epsilon \nabla^2 \Phi = \sum_i e^2 z_i^2 \rho_{si} \Phi / k_B T. \quad (2)$$

For instance, by implementing effective charges instead of real charges, the dressed-ion theory [20] yields a Yukawa potential as the solution to the above DH equation, and the molecular Debye-Hückel theory [21, 22] utilizes a linear combination of Yukawa potentials. However, methods of this type have the drawback that the dielectric function of the system must be known in advance to determine key parameters.

Recently, the field-theoretic (FT) approach provides an elegant way of solving the grand partition function, leading to a correction term named the *self-energy* [23]. Below we confine ourselves to the case of having only two ion species whose charge valances are $z_+$ and $z_-$, then the self-energy in the FT approach at position $r$ can be defined as:

$$u_\pm^{FT}(r) = \frac{1}{2} z_\pm^2 \lim_{r' \to r} \left[ G(r,r') - \frac{l_B}{|r-r'|} \right], \quad (3)$$

where $G(r,r')$ is a Green's function to be determined and $l_B \equiv \frac{e^2}{4\pi\epsilon k_B T}$ is the Bjerrum length. Interestingly, several different methods, such as the variational method [23-27] and the loop-wise expansion [28], end up with the same expression for $G(r,r')$:

$$-\nabla^2 G(r,r') + 4\pi l_B \left( z_+^2 \rho_+^{FT} + z_-^2 \rho_-^{FT} \right) G(r,r')$$
$$= 4\pi l_B \delta(r-r'), \quad (4)$$

and correspondingly the PB equation is modified as

$$-\nabla^2 \phi = 4\pi l_B \left( \lambda_+ z_+ e^{-z_+\phi - u_+^{FT}} + \lambda_- z_- e^{-z_-\phi - u_-^{FT}} + \rho_{ex} \right), \quad (5)$$

where the dimensionless potential $\phi \equiv \frac{e}{k_B T} \Phi$ is used for simplicity, and a fixed ion distribution $\rho_{ex}$ is included for generality. The ion density is $\rho_\pm^{FT} = \lambda_\pm e^{-z_\pm\phi - u_\pm^{FT}}$, where $\lambda_\pm = e^{\mu_\pm}/v_\pm$ is the fugacity of ions with $\mu_\pm$ being the chemical potential and $v_\pm$ being a volume scale in the partition function [23]. Note that if the ion density $\rho_\pm^{FT}$ is a constant, then Eq. (4) reduces to the original DH equation. We therefore call it a DH-like equation for the Green's function. Although the DH-like equation Eq. (4) looks linear, it is actually non-linear because it depends implicitly on the self-energy. However, better accuracy can be expected by replacing the DH-like equation by a certain PB-like equation for the Green's function, since the original DH equation is just a linear approximation of the original PB equation [29-31]. On the other hand, numerical solutions to Eqs. (3)-(5) are intensively studied [32-34], but no molecular simulations have been performed to directly justify the accuracy of those methods. As we will show below by our molecular dynamics (MD) simulation, this DH-like equation provides good corrections only for the counter-ion distribution, and fails for the co-ion distributions. To incorporate the ion correlation effect more accurately into the original PB theory, we herein present a modified PB theory with non-linear ion correlations, whose self-energy is reinterpreted with a PB-like equation from a physics point of view. As justified by our MD simulation results, our new equation provides excellent results for both counter-ion and co-ion distributions, and thus can be regarded as successful for both the self-energy and the correction for the original PB equation.

This paper is organized as follows. In section 2, we show the derivation of our equation. In section 3, we propose our model and show all the results from both numerical solutions and MD simulations. The problems of the FT approach as well as the limitations of our theory are discussed in section 4. Finally we summarize the key ideas of this work in section 5.

## 2. Methods

To find a modified PB equation incorporating the self-energy, we start from the original two-species PB equation with the dimensionless potential

$$-\nabla^2 \phi_{PB} = 4\pi l_B \left( z_+ \rho_{s+} e^{-z_+\phi_{PB}} + z_- \rho_{s-} e^{-z_-\phi_{PB}} + \rho_{ex} \right), \quad (6)$$

where the subscript "PB" stands for the solution to the original PB equation, $\rho_{s+}$ and $\rho_{s-}$ are average densities of the two ion species. To add in ion correlations, we now insert a test ion at position $r'$, whose bare potential $G_0(r,r')$ at position $r$ is given by the Green's function as

$$-\nabla^2 G_{0\pm}(r,r') = 4\pi l_B z_\pm \delta(r-r'). \quad (7)$$

When the system is fully relaxed to a new equilibrium state after the insertion of the test ion, the average potential is perturbed to be $\phi_{PB}(r) + G_\pm(r,r')$, where $G_\pm(r,r')$ is the incremental potential due to the test ion. The corresponding equation for such a system is

$$-\nabla^2 \left[ \phi_{PB}(r) + G_\pm(r,r') \right] = 4\pi l_B [z_+\rho_{s+} e^{-z_+\phi_{PB}(r) - z_+ G_\pm(r,r')}$$
$$+ z_-\rho_{s-} e^{-z_-\phi_{PB}(r) - z_- G_\pm(r,r')} + \rho_{ex} + z_\pm \delta(r-r')]. \quad (8)$$

In contrast to the original PB equation Eq. (6), we find the following equation for $G_\pm(r,r')$ by a simple subtraction:

$$-\nabla^2 G_\pm(r,r') = 4\pi l_B [z_+\rho_{s+} e^{-z_+\phi_{PB}(r)} \left( e^{-z_+ G_\pm(r,r')} - 1 \right)$$
$$+ z_-\rho_{s-} e^{-z_-\phi_{PB}(r)} \left( e^{-z_- G_\pm(r,r')} - 1 \right) + z_\pm \delta(r-r')]. \quad (9)$$

Eq. (9) indicates that two factors contribute to $G_\pm(r,r')$: the presence of the test ion and the change of surrounding ions due to the presence of the test ion. The potential induced by the change of surrounding ions results from ion correlations:

$$u_\pm(r') = \lim_{r \to r'} u_\pm(r, r') = \lim_{r \to r'} [G_\pm(r, r') - G_{0\pm}(r, r')], \quad (10)$$

which is exactly the self-energy by definition. With the correction of the self-energy, the ion densities become

$$\rho_\pm(r) = \rho_{0\pm} e^{-z_\pm \phi(r) - z_\pm u_\pm(r)}, \quad (11)$$

where $\phi$ is the new electric potential to be solved, $\rho_{0\pm} = \rho_{s\pm} e^{z_\pm u_\infty}$ with $u_\infty$ being the self-energy at infinity to ensure that $\rho_\pm(r)$ approaches $\rho_{s\pm}$ when $r$ goes to infinity. The PB equation is then modified to be

$$-\nabla^2 \phi = 4\pi l_B \left( z_+ \rho_{0+} e^{-z_+ \phi - z_+ u_+} + z_- \rho_{0-} e^{-z_- \phi - z_- u_-} + \rho_{ex} \right), \quad (12)$$

and the corresponding Green's function is determined by

$$-\nabla^2 G_\pm(r, r') = 4\pi l_B [z_+ \rho_+(r) \left( e^{-z_+ G_\pm(r, r')} - 1 \right) + z_- \rho_-(r) \left( e^{-z_- G_\pm(r, r')} - 1 \right) + z_\pm \delta(r - r')]. \quad (13)$$

Note that now Eq. (13) is different from Eq. (9) by the Boltzmann factor $-z_\pm \phi_{PB} \to -z_\pm \phi(r) - z_\pm u_\pm(r)$. This is the PB-like equation for Green's function. As expected, our theory reduces to the results of the FT approach if one applies a linear approximation, whose derivations are shown in the Appendix.

However, such a linearization is in fact mathematically problematic since $G_\pm(r, r')$ diverges as $r$ goes to $r'$, so the linearization is acceptable only when $G_\pm(r, r')$ is close to zero. As a result, our theory and the FT approach should lead to qualitatively different results.

## 3. Results

### 3.1 Model

In this section we justify the accuracies of both our theory and the FT approach by MD simulation.

To avoid the divergence problem due to the singularity in electrostatic interactions, we treat each ion as a soft ball instead of a point charge by introducing between ions the van der Waals (VDW) interaction as

$$v(r) = \frac{4\varepsilon}{k_B T} \left[ \left(\frac{\sigma}{r}\right)^{12} - \left(\frac{\sigma}{r}\right)^6 \right], \quad (14)$$

where $v(r)$ is the Lennard-Jones (LJ) potential rescaled by a factor of $1/k_B T$, $\varepsilon$ is the depth of the potential well and $\sigma$ is the distance at which the LJ potential is zero. For the sake of simplicity, we assume that both ion species have the same VDW parameters, and the VDW interactions are also treated in a mean-field way. Consequently, by subtracting Eq. (7) from Eq. (13) and including the VDW interactions, the *self-energy equation* becomes

$$-\nabla^2 u_\pm(r, r') = 4\pi l_B [z_+ \rho_+(r) \left( e^{-z_+ [u_\pm(r, r') + G_{0\pm}(r, r')] - v(|r - r'|)} - 1 \right) + z_- \rho_-(r) \left( e^{-z_- [u_\pm(r, r') + G_{0\pm}(r, r')] - v(|r - r'|)} - 1 \right)]. \quad (15)$$

In general, Eq. (15) can only be solved numerically. Eq. (7) is also solved numerically by replacing the Dirac delta function by the Kronecker delta function to avoid divergence. The boundary condition for Eq. (15) is

$$\lim_{r \to \infty} u(r, r') = \frac{l_B}{|r - r'|} \left( e^{-\kappa_s |r - r'|} - 1 \right)$$

since Eq. (13) should reduce to the original DH equation at large $r$. With the help of Eq. (7), Eq. (15) can be iteratively solved to determine $u_\pm(r, r')$ by employing the Finite-Different method, and then $u_\pm(r)$ can be obtained by taking the limit $r \to r'$.

The structure of the ionic systems is often described by the radial distribution function (RDF), which is the density distribution with respect to a reference particle and can be directly calculated in MD simulations. In order to calculate the RDF by the modified PB theories, we add a fixed ion as the reference particle, which is equal to set $\rho_{ex} = z_\alpha \delta(r)$ in Eqs. (5) and (12). Then the RDF is calculated as

$$g_{\alpha\pm}(r) = e^{-z_\pm \phi(r) - z_\pm u_\pm(r) - v(r)}. \quad (16)$$

To find the solutions, one can start with $u(r) = 0$, and then solve Eqs. (12) and (15) iteratively until the solutions converge. In this work, convergence is reached after 3 iterations, except for the FT approach in the charge-asymmetric case, where the solution diverges with iteration. Therefore, for the FT approach in the charge-asymmetric case, the solution after one iteration is shown, while for other cases the solutions after three iterations are shown.

### 3.2 Numerical methods

Because this system is isotropic with respect to $r = 0$, the location of the reference ion, the potential $\phi(r)$ and the self-energy $u(r)$ become one-dimensional in the spherical coordinates.

The self-energy function $u(r, r')$ is not spherically symmetric with respect to $r = 0$, but it is symmetric with respect to the line connecting the two points $r = 0$ and $r = r'$. The self-energy equation is then reduced to two dimensional in cylindrical coordinate and we have $u(s, \theta, z) = u(s, z)$, where $z$ is the direction from $r = 0$ to $r = r'$, $s$ is the distance from z-axis, and $u$ is independent of $\theta$ due to azimuthal symmetry. By discretizing Eq. (15), we obtain

$$-u_{ss} - \frac{1}{s} u_s - u_{zz} = 4\pi l_B [z_+ \rho_+(r) \left( e^{-z_+ [u_\pm(r, r') + G_{0\pm}(r, r')] - v(|r - r'|)} - 1 \right) + z_- \rho_-(r) \left( e^{-z_- [u_\pm(r, r') + G_{0\pm}(r, r')] - v(|r - r'|)} - 1 \right)]. \quad (17)$$

In order to address the divergence $\lim_{r \to 0} G_0(r) = \lim_{r \to 0} \frac{l_B}{|r|}$, equivalent to replacing the Dirac delta by the Kronecker delta, $G_0(r)$ is approximated by the solution of a uniformly charged sphere of diameter $h$, which is the finite difference interval, instead of a point charge, to obtain

$$G_0(r) = \begin{cases} \dfrac{l_B}{r}, & r > \dfrac{h}{2} \\ \dfrac{l_B}{h}\left(3 - \dfrac{4r^2}{h^2}\right), & r < \dfrac{h}{2} \end{cases}. \quad (18)$$

Thus we have $G_0(0) = \dfrac{3l_B}{h}$. Using centered difference method we have:

$$u_s(s,z) = \frac{u(s+h,z) - u(s-h,z)}{2h} + O(h^2), \quad (19)$$

$$u_{ss} = \frac{u(s+h,z) - 2u(s,z) + u(s-h,z)}{h^2} + O(h^2), \quad (20)$$

$$u_{zz} = \frac{u(s,z+h) - 2u(s,z) + u(s,z-h)}{h^2} + O(h^2). \quad (21)$$

For the divergence at $s = 0$, one can either use the Cartesian coordinate at this point, or use the L'Hospital's rule to get

$$\nabla^2 u(0,z) = \frac{4u(s,h) + u(s+h,0) + u(s-h,0) - 6u(s,0)}{h^2}. \quad (22)$$

After discretization, Eq. (17) is solved by using the Successive Over-Relaxation (SOR) method.

### 3.3 MD simulations

To compare with the modified PB theories which treat the solvents implicitly, an implicit-solvent aqueous ionic solution is modelled in the simulations. A cubic box with a side length of 20 nm is used in the simulations. Both charge-symmetric and charge-asymmetric cases are simulated. For the charge-symmetric case, 1920 monovalent ion pairs are put in the box, resulting in a density of 0.4 M. For the charge-asymmetric case, 1920 monovalent ions and 960 divalent counter-ions are put in the same simulation box. The periodic boundary conditions are applied in all three dimensions.

Water molecules are not explicitly presented in the simulation setup since the PB theories imply a continuous solvent. In order to allow the system to have $\epsilon_w \approx 80$, the dielectric constant of water, the electrostatic force between two ions is rescaled by a factor of $1/\epsilon_w$. In our simulations, all the ion charges are rescaled by a factor of 0.11, corresponding to rescaling the electrostatic force by a factor of 1/82.6, which results in a dielectric constant of 82.6. The Particle-Mesh Ewald method is used to deal with electrostatic interactions. The VDW interactions between ions have the same force-field parameters $\varepsilon = 0.0696357$ kJ/mol and $\sigma = 0.386472$ nm. The simulation runs for 10 ns at $T = 330$ K with a time step of 2 fs and a sampling interval of 0.2 ps. The simulation is performed with the Gromacs MD simulation package [35].

### 3.4 Charge-symmetric case

In the charge-symmetric case, a positive ion is fixed as the reference particle. The self-energy $u_\pm$ calculated by our theory (Eqs. (10) and (15)) and $u_\pm^{FT}$ calculated by the FT approach (Eqs. (3) and (4)) are compared in figure 1. According to Eq. (3), in the FT approach all the valance terms are squared, $z_+^2 = z_-^2$, so there are no differences in calculating the self-energies $u_-^{FT}$ for counter-ions and $u_+^{FT}$ for co-ions, leading to the same self-energy curves for counter-ions and co-ions. Such a result is physically questionable since the self-energy depends on the distributions of counter-ion and co-ion with respect to the reference ion, which obviously should be different. On the contrary, the self-energy curves in our theory are different for counter-ions and co-ions. The questionable squared valence in the FT approach is the result of the linearization $z_\pm\left(e^{-z_\pm G_\pm} - 1\right) \approx -z_\pm^2 G_\pm$ as described in section 2.

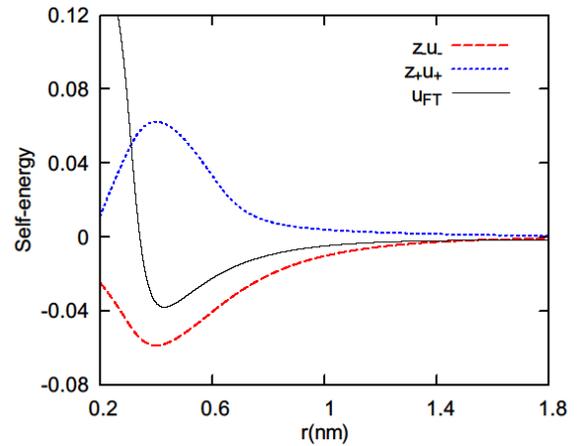

**Figure 1.** The self-energy of counter-ions $z_- u_-$ (red) and co-ions $z_+ u_+$ (blue) calculated by our theory (Eqs. (10) and (15)) in the charge-symmetric case. The self-energies $u_\pm^{FT}$ (black) of counter-ions and co-ions in the FT approach are the same in the charge symmetric case according to the definition in Eq. (3). Note that the self-energy in this work should be multiplied by the charge valency to compare with the FT self-energy (see the Appendix for detail).

The RDFs obtained from the MD simulation are compared with those calculated according to the original PB theory, the FT approach, and our theory in figure 2. The counter-ion and co-ion RDFs are denoted as $g_{+-}(r)$ and $g_{++}(r)$, respectively. It is clear that the original PB theory underestimates the counter-ion RDF peak height and overestimates the co-ion RDF peak height. For the counter-ion RDFs, both modified theories make clear corrections to the PB theory, while our theory obviously yields better results, as demonstrated by the figures. For the co-ion RDFs, our theory still yields better results than the original PB theory, but the FT approach leads to an even worse co-ion RDF than the original PB theory, attributed to the fact that the DH-like equation of the FT approach has identical self-energies for positive and negative ions, according to Eqs. (3) and (4).

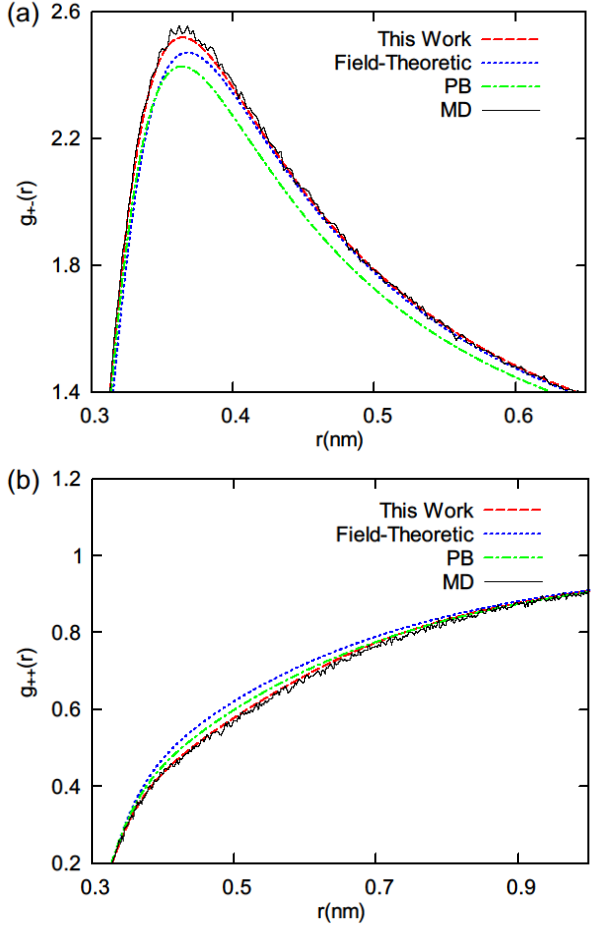

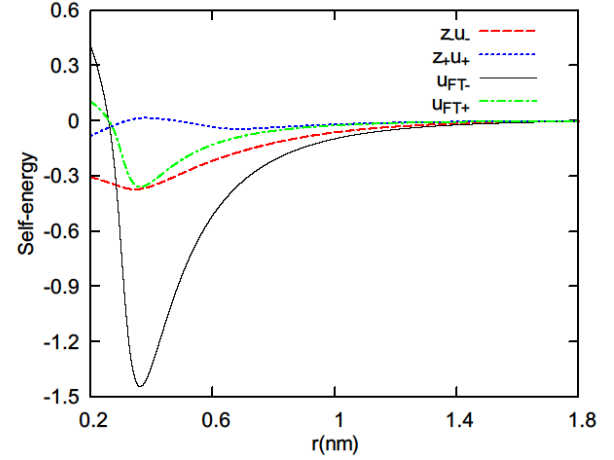

**Figure 3.** The self-energy $z_\pm u_\pm$ calculated by of our theory (Eqs. (10) and (15)) and $u_\pm^{FT}$ in the FT approach (Eqs. (3) and (4)) in the charge-asymmetric case. There are notable differences between the co-ion self-energies. Note that the self-energy in this work should multiply the charge valency to compare with the FT self-energy (see the Appendix for detail).

counter divalent ion distributions should be 4 times larger than $u_+^{FT}$ according to the definition.

As shown in figure 4, the RDFs for both counter-ion and co-ion calculated by our theory still match the MD simulation very well, while those calculated by the FT approach deviate from MD results significantly, indicating the FT approach fails in this case. Moreover, the FT approach is in fact divergent in this case, since the height of RDFs will be higher and higher after iterations, as mentioned in Section 3.1.

**Figure 2.** Comparison of RDFs between the theories and the MD simulations (black) in the charge-symmetric case. The theoretical RDFs are calculated by Eq. (16) with the original PB theory (green), the FT approach (blue), and our theory (red), and a positive ion is used as the reference particle. (a) Counter-ion RDFs. (b) Co-ion RDFs.

*3.5 Charge-asymmetric case*

The self-energies and RDFs of the charge-asymmetric ion system are also calculated by the theories and compared with MD simulations. In this case we use a positive monovalent ion as the reference particle and then calculate the ion distributions. The self-energy curves calculated by our theory and by the FT approach are shown in figures 3(a) and 3(b), respectively. In this case there are notable differences between the self-energies of our theory and the FT approach, especially for the result of $u_-^{FT}$. In the FT approach, $u_-^{FT}$ corresponding to the modification for the

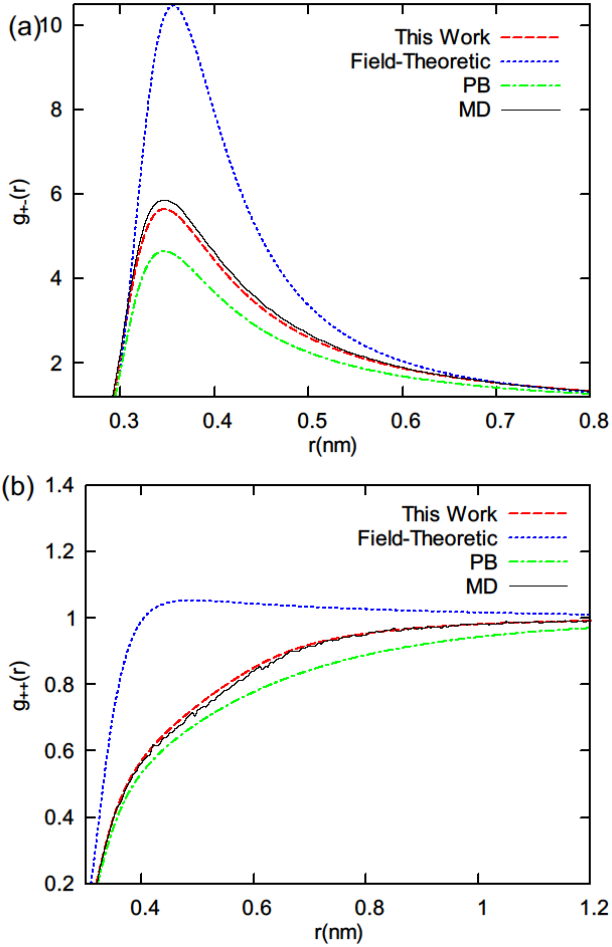

**Figure 4.** Comparison of RDFs between the theories and the MD simulations (black) in the charge-asymmetric case. The theoretical RDFs are calculated by Eq. (16) with the original PB theory (green), the FT approach (blue) and our theory (red), and a positive monovalent ion is used as the reference particle. (a) Counter-ion RDFs. (b) Co-ion RDFs.

## 4. Discussion

The comparisons in figures 2 and 4 indicate a remarkable degree of consistency between the MD simulation and our theory for both counter-ion and co-ion RDFs, while the FT approach is only good for the counter-ion RDF and fails for the co-ion RDF. The failure can be attributed to the *z*-squared terms in the DH-like equation for the self-energy. The *z*-squared terms appear when our PB-like self-energy equation is linearized, but this kind of linearization is mathematically problematic. The FT approach is based on a variational method requiring an undetermined reference action to perform the derivations. The most commonly used reference action has the Gaussian form, but the validity of this Gaussian assumption is not easy to justify [36]. Therefore, the problem of the DH-like equation very likely comes from the questionable assumption of the Gaussian reference action.

Although our theory predicts more accurate RDFs than the FT approach, we note that there are still small deviations from the MD simulations, which may be attributed to two factors. The first one is the mean-field approximation for the VDW interactions. To test if the non-electrostatic interaction plays an important role, a benchmark run is performed by turning off the ion charges and comparing this simple approach with the MD simulation. The results given in figure 5 show that the RDF difference between the theory and the MD simulation in the charge-free case is no larger than 0.01 at each point, which is negligible compared with the ionic case, so the approximation we used for the non-electrostatic interaction is not a major source of error. The second one is that the higher-order electrostatic correlation effects are not considered in our theory. We employ the same approach for calculating RDFs theoretically, whose results are good enough for drawing the conclusions clearly. Since it is not the goal of this work to calculate RDFs very accurately, these approximations should not influence our qualitative conclusions.

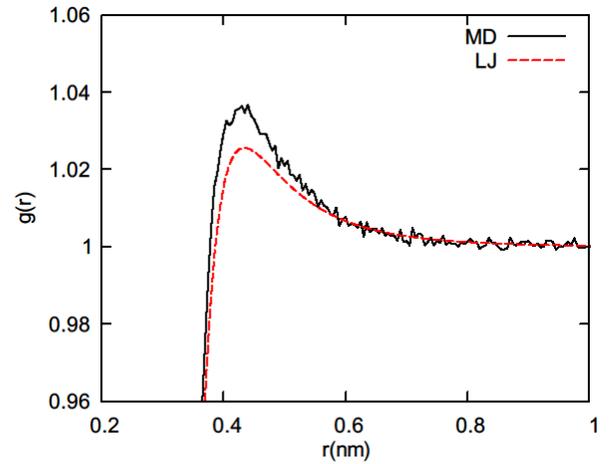

**Figure 5.** The RDFs of a simple LJ particle system obtained by the LJ potential and the MD simulation (see the text for details).

In the case of calculating the RDF for a homogeneous system, our modified PB theory only provides quantitative corrections to the original PB theory. However, the concept of self-energy is important in many applications. For example, in the electric double layer problem, the image charge due to dielectric discontinuity can be mathematically described by the self-energy. The image charge interaction can be dominant at the boundary of dielectric discontinuity, leading to qualitatively different results from the original PB equation even in the weak-coupling limit [32, 34, 37]. Our model for the self-energy is also helpful for the studies of other correlation-induced phenomena, such as charge inversion and attraction between like-charged colloids in ionic solutions. Compared to numerous existing works trying to modify the original PB equation, one of the key advantages of our theory is the equations are easy to be applied to practical situations. Besides the extensively studied double-layer or colloid systems, our PB-like self-energy equation can be directly solved numerically without further approximations for charged systems with an arbitrary structure.

## 5. Conclusions

In this work, we introduce a self-energy function from a physics point of view to obtain a novel modified PB theory by utilizing the Green's function. The ion correlation effect is successfully described by the self-energy. Our modified PB theory can be reduced to the FT approach by applying a linear approximation. MD simulations are performed to test the validity of the theories by comparing the RDFs of an implicit-solvent ionic system. Both theories make clear corrections to the original PB theory for the counter-ion RDFs, and our theory provides better results as expected. The FT approach fails for the co-ion RDFs, while our theory still works and is better than the original PB theory. The failure of the FT approach may be attributed to the *ad hoc* Gaussian assumption for the reference action in its derivations. Although we have only tested the validity of our modified PB theory with a homogeneous two-species ionic solution, it is apparently applicable to ionic systems with multiple ion species and arbitrary boundary conditions. Our theory is hopefully to be applied to the study of correlation-significant systems, such as double-layer structure, charge inversion phenomena and attraction between like-charged colloids in ionic solutions.

## Acknowledgements

This work was funded by the National Natural Science Foundation of China (Nos. 11774357 and 11747601). Y. W. also thanks the financial support through the CAS Biophysics Interdisciplinary Innovation Team Project (No. 2060299). Allocations of computer time from the Constance Cluster of PNNL and the HPC cluster of ITP-CAS are gratefully acknowledged.

## Appendix. Derivation from our theory to the FT approach by linearization

Starting from the PB-like equation for the Green's function (Eq. (13) in the main text):

$$-\nabla^2 G_\pm(r,r') = 4\pi l_B [z_+\rho_+(r)\left(e^{-z_+ G_\pm(r,r')}-1\right) \\ + z_-\rho_-(r)\left(e^{-z_- G_\pm(r,r')}-1\right) + z_\pm\delta(r-r')]. \quad (A1)$$

The linear approximation $e^{-z_+ G_\pm(r,r')} - 1 \approx -z_+ G_\pm(r,r')$ leads to:

$$-\nabla^2 G_\pm(r,r') + 4\pi l_B \left(z_+^2\rho_+(r)+z_-^2\rho_-(r)\right)G_\pm(r,r') \\ = 4\pi l_B z_\pm \delta(r-r'). \quad (A2)$$

The field-theoretic (FT) Green's function is

$$-\nabla^2 G(r,r') + 4\pi l_B \left(z_+^2\rho_+^{FT}(r)+z_-^2\rho_-^{FT}(r)\right)G(r,r') \\ = 4\pi l_B \delta(r-r'). \quad (A3)$$

Note that the coefficients of $\delta(r-r')$ are $4\pi l_B z_\pm$ in Eq. (A2) and $4\pi l_B$ in Eq. (A3). Therefore, $G_\pm(r,r')$ in the PB-like equation (our theory) should correspond to $z_\pm G(r,r')$ in the DH-like equation (the FT approach). Define the self-energy:

$$u_\pm(r') = \lim_{r \to r'}\left[G_\pm(r,r') - G_{0\pm}(r,r')\right], \quad (A4)$$

where $G_{0\pm}(r,r')$ is defined in Eq. (7) in the main text, and

$$u_\pm^{FT}(r) = \frac{1}{2}z_\pm^2 \lim_{r' \to r}\left[G(r,r') - \frac{l_B}{|r-r'|}\right]. \quad (A5)$$

We present all physical quantities as functions of $r$ and $r'$ with the latter being a constant representing the location of the test ion. Since $G_\pm(r,r')$ and $G_{0\pm}(r,r')$ are the potentials at $r$ induced by the ion at $r'$, the self-energy at $r'$ can be obtained by taking the limit $r \to r'$, and then $u(r')$ is replaced by $u(r)$ when solving the modified PB equation.

The coefficient in the definitions of the self-energy and the coefficients of u in the modified PB equations are different in our theory and in the FT theory, whose modified PB equations are

$$-\nabla^2\phi = 4\pi l_B \left(z_+\rho_{0+}e^{-z_+\phi-z_+ u_+} + z_-\rho_{0-}e^{-z_-\phi-z_- u_-}\right), \quad (A6)$$

and

$$-\nabla^2\phi = 4\pi l_B \left(\lambda_+ z_+ e^{-z_+\phi-u_+^{FT}} + \lambda_- z_- e^{-z_-\phi-u_-^{FT}}\right), \quad (A7)$$

respectively, where $\rho_{0\pm}$ and $\lambda_\pm$ are the same as described in the main text and the references therein. The terms $z_\pm u_\pm$ in Eq. (A6) and $u_\pm^{FT}$ in Eq. (A7) are the corrections (self-energy) to the original PB equation. The self-energy comparisons are shown in FIGs. 1 and 3.

Now it is easy to prove that, under the linear approximation, the equation set (A2)(A4)(A6) of our nonlinear PB theory is equal to the equation set (A3)(A5)(A7) of the FT approach, except for a factor of 1/2 in the definitions of the self-energy.

The major difference between our equations and the FT equations is the treatment of the valances $z_\pm$. In the FT equations, they always appear as $z_+^2$ and $z_-^2$, so the self-energies $u_+$ and $u_-$ are the same for a symmetric electrolyte which is unphysical and inconsistent with our MD simulation results. Moreover, the linearization $e^{-z_{\pm}G_{\pm}(r,r')} - 1 \approx -z_{\pm}G_{\pm}(r,r')$ is in fact mathematically problematic since $G(r,r')$ diverges as $r$ goes to $r'$, which is acceptable only when $G(r,r')$ is close to zero.